\documentclass[showpacs,a4paper,prc,aps,twocolumn]{revtex4}
\usepackage[dvips]{graphicx}
\usepackage{color}
\usepackage{amssymb}
\usepackage{amsmath}
\usepackage{amsfonts}
\usepackage{indentfirst}

\begin{document}

\title{Decay Widths of X(1835) as $\overline NN$ Bound State}
\author{D. Samart\footnotemark[1], Y. Yan\footnotemark[1]\footnotemark[2]\footnotemark[4], Th. Gutsche\footnotemark[3], Amand Faessler\footnotemark[3]
\vspace*{0.4\baselineskip}}
\affiliation{
\footnotemark[1]
School of Physics, Suranaree University of Technology, \\
111 University Avenue, Nakhon Ratchasima 30000, Thailand \\
\footnotemark[2]
ThEP Center, Commission on Higher Education, Bangkok 10400, Thailand \\
\footnotemark[3]
Institut f\"ur Theoretische Physik, Universit\"at T\"ubingen, Auf der Morgenstelle 14, D-72076 T\"ubingen, Germany \\
\footnotemark[4]
Email: yupeng@sut.ac.th
\vspace*{0.4\baselineskip}}

\begin{abstract}
Partial decay widths of various decay channels of the $X(1835)$ are evaluated in the $^3P_0$ quark model,
assuming that the $X(1835)$ is a $N\overline N$ bound state with the quantum number
assignment $I^G(J^{PC})=0^+(0^{-+})$. It is found that the decays to the $\rho\rho$, $\omega\omega$ and $\pi a_0(1450)$
states dominate over other channels, and that the product branching fractions
$Br(J/\psi\to\gamma X)Br(X\to \pi\pi\eta)$ and $Br(J/\psi\to\gamma X)Br(X\to \pi\pi\eta')$
are in the same order. We suggest that the $X(1835)$ may be searched in the $\pi a_0(1450)$ channel.

\end{abstract}

\maketitle

\section{Introduction}
An enhancement was observed by the BES Collaboration \cite{BES2003}
in the proton-antiproton ($p\overline p$) invariant mass spectrum in the radiative
decay $J/\psi\to\gamma p\overline p$.
It was concluded that the enhancement has properties consistent with either a $J^{PC}=0^{-+}$ or $0^{++}$
quantum number assignment and unlikely stems from the effects
of any known meson resonance. The mass and width of the
resonance are fixed to be $M=1859^{+3}_{-10}({\rm stat})^{+5}_{-25}({\rm syst})$ MeV and $\Gamma <30$ MeV
if it is interpreted as a single $J^{PC}=0^{-+}$ resonance.
More recently the BES Collaboration \cite{BES2005} analyzed the decay channel $J/\psi\to\gamma\pi^+\pi^-\eta'$ and
observed a resonance, the $X(1835)$ with high statistics in the $\pi^+\pi^-\eta'$ invariant mass spectrum,
with the product branching fraction
$Br(J/\psi\to\gamma X(1835))Br(X(1835)\to\pi^+\pi^-\eta') = (2.2\pm 0.4({\rm stat.})\pm 0.4({\rm syst.}))\times 10^{-4}$.
The mass and width of the $X(1835)$ are determined to be $1833.7\pm 6.1 ({\rm stat})\pm 2.7( {\rm syst})$
MeV and $67.7\pm 20.3({\rm stat}) \pm 7.7( {\rm syst})$ MeV, respectively.
The possibility that the X(1835) and the resonance reported in Ref. \cite{BES2003} are the same
entity was checked in Ref. \cite{BES2005}. Redoing the S-wave Breit-Wigner fit to the $p\overline p$
invariant mass spectrum of Ref. \cite{BES2003}, including the effect of
final-state-interactions on the shape of the $p\overline p$ mass
spectrum \cite{Zou2003,Sibirtsev2005}, yields a mass
$M =1831\pm 7$ MeV and a width $\Gamma < 153$ MeV which are consistent with the
$X(1835)$ observables in Ref. \cite{BES2005}. The $X(1835)$ was confirmed by the BESIII experiment \cite{BES2011}
with a statistical significance larger than $20\sigma$. Up to now, however, the spin and parity of the $X(1835)$
has not been well determined.

The nature of the $X(1835)$ is still an open question though numbers of theoretical works have
been done to interpret this particle. Among the interpretations are the $\overline NN$ bound state
\cite{Loiseau1,Loiseau2,ppbar1,ppbar2,ppbar3,ppbar4,ppbar5,ppbar6,ppbar7,ppbar8,ppbar9},
the baryonium with sizable gluon content \cite{ppbarG},
the pseudoscalar glueball \cite{gluonball1,gluonball2,gluonball3,gluonball4},
the radial excitation of the $\eta'$ \cite{etap1,etap2,etap3,etap4},
and the $\eta_c$-glueball mixture \cite{gluonballC1,gluonballC2}.
The $\overline NN$ bound state interpretation has been the most natural one
since the $X(1835)$ resonance is a prime candidate for the source of the $p\overline p$
invariant mass enhancement in $J/\psi\to\gamma p\overline p$ reaction.
However, the fact that the $X(1835)$ resonance is not observed
in the $\pi^+\pi^-\eta$ invariant mass spectrum has been an outstanding challenge to
the $N\overline N$ bound state interpretation.

The BESII measurement \cite{BES2003} of the photon polar angle
distribution in radiative $J/\psi$ decays favors a $J=0$ $p\overline p$ system,
but the possibility of a $J=1$ resonance is not excluded. However, the recent BESIII
experiment \cite{BES2011} reveals that the
polar angle of the photon in the process $J/\psi\to\gamma X(1835)$ agrees well
with the form $1 + \cos\theta_\gamma$, which indicates that the $X(1835)$ is either
a scalar or pseudoscalar meson.
Therefore, interpreting the $X(1835)$ as an $N\overline N$ bound state means that
the $X(1835)$ could be a $^{11}S_0$, $^{31}S_0$, $^{13}P_0$, or $^{33}P_0$ state.

The investigations in Ref. \cite{Maruyama:1987tx,Maruyama:1987kt,Dover:1990}
of $N\overline N$ annihilations and $N\overline N$ bound states in the $^{3}P_0$
quark model reveal that the dominant decay modes of $^{31}S_0$, $^{11}S_0$, $^{13}P_0$ and $^{33}P_0$
$N\overline N$ bound states
are the $\rho\omega$ and $\pi\rho$ channels, the $\rho\rho$ and $\omega\omega$ channels,
the $\pi\pi$, $\eta\eta$, $\rho\rho$ and $\omega\omega$ channels and the
$\pi\eta$ and $\rho\omega$ channels, respectively.
It may be difficult for the BES detectors to
observe resonances in the $4\pi$, $5\pi$ or $6\pi$ channel, but we expect that
the BES Collaboration is able to retrieve a resonance if it decays mainly to the $\pi\pi$,
$3\pi$, $\eta\eta$
or $\pi\eta$ state. Up to now there has been no report from the BES Collaboration of
such a resonance from the $\pi\pi$,
$3\pi$, $\eta\eta$ or $\pi\eta$ channel. One may conclude that the $X(1835)$ is unlikely to be a
$^{31}S_0$, $^{13}P_0$ or $^{33}P_0$ $N\overline N$ bound state, and hence that a $^{11}S_0$
$N\overline N$ bound state might be the only candidate for the $X(1835)$. Studies of
the $J/\psi$ decays
$J/\psi\to \gamma\pi^+\pi^-\eta$ and $\gamma\,\overline pp$ in Ref. \cite{Loiseau1,Loiseau2} in
a semi-phenomenological potential model reveal that the explanation of both the reactions may be
given by a broad $^{11}S_0$ $\overline NN$ bound state near the $\overline NN$ threshold. However,
the investigations of the spectrum of $\overline NN$ bound states in microscopically derived $\overline NN$
potentials \cite{Yan1997} and in phenomenological potentials \cite{Dover:1990} have not
found such a $^{11}S_0$ $\overline NN$ bound state.

In this work we evaluate the partial decay widths for various decay channels of the $X(1835)$
in the $^3P_0$ quark model, assuming that
the $X(1835)$ is a $^{11}S_0$ $N\overline N$ bound state with the quantum number
assignment $I^G(J^{PC})=0^+(0^{-+})$. The paper is arranged as follows: In Section II we calculate
the decay widths for two-body decay channels. The partial decay widths for the $\pi^+\pi^-\eta$ and
$\pi^+\pi^-\eta'$ channels are estimated in Section III. Discussion and conclusions are
given in Section IV.

\section{$X(1835)$ to two mesons}
Interpreted as a $N\overline N$ bound state with the quantum number
assignment $I^G(J^{PC})=0^+(0^{-+})$, the $X(1835)$ may mainly decay to the two-meson final states
$\rho\rho$, $\omega\omega$, $\pi a_0(1450)$, $\eta f_0(1370)$, $\eta' f_0(1370)$,
$\pi a_2(1230)$ and $\eta f_2(1270)$ as well as $\pi a_0(980)$, $\eta f_0(600)$ and $\eta' f_0(600)$.
The transition amplitude of $X(1835)$ to two mesons takes the form
\begin{eqnarray}\label{eq::1}
    T_{X\to M_1M_2}&=& \langle M_1M_2|V|\overline NN\rangle\langle \overline NN|X\rangle \nonumber \\
    &=& \int\frac{d\vec
    p}{(2\pi)^{3/2}}~\Phi_X(\vec p\,)~T_{\overline NN \to M_1M_2}(\vec p,\vec k\,)
\end{eqnarray}
where $\Phi_X(\vec p)$ is the $N\overline N$ bound state wave function of the $X(1835)$
in momentum space, normalized according to
\begin{eqnarray}
\int\frac{d\vec p\,}{(2\pi)^3}|\Phi_X(\vec p\,)|^2=1,
\end{eqnarray}
$\vec k$ is the relative momentum between the two final mesons, and $\vec p$
the relative momentum between the nucleon and antinucleon of the $X(1835)$ in the
center-of-mass system. $T_{\overline NN \to M_1M_2}(\vec p,\vec k)$ is the transition amplitude
of a free nucleon-antinucleon pair to two mesons.

In this work we study $N\overline N$ annihilations to two mesons in
the A2 quark line model, as described in \cite{Dover:1992vj}, where the effective quark
annihilation operator takes the quantum numbers of the vacuum ($^3P_0$, isospin $I=0$ and
color singlet). Meson and baryon decays and $N\bar N$ annihilation into two mesons
are well described phenomenologically using such an effective
quark-antiquark vertex. At least from meson decays, this approximation has been
given a rigorous basis in strong-coupling QCD. The nonperturbative
$q\bar{q}$ $^3P_0$ vertex is defined according to Ref.
\cite{Dover:1992vj}
\begin{eqnarray}
V^{ij}=\lambda\,\sum_\mu\sigma^{ij}_{-\mu}Y_{1\mu}(\vec{q}_i-\vec{q}_{j})\delta^{(3)}(\vec{q}_i+\vec{q}_{j})(-1)^{1+\mu}1^{ij}_F1^{ij}_C~,
\end{eqnarray}
where $\lambda$ is the effective coupling strength of the $^3P_0$ vertex,
$Y_{1\mu}(\vec{q}\,)=|\vec{q}\,|\mathcal{Y}_{1\mu}(\widehat{q})$
with $\mathcal{Y}_{1\mu}(\widehat{q})$ being the spherical harmonics
in momentum space, and $1^{ij}_{F}$ and $1^{ij}_{C}$ are unit
operators in flavor and color spaces, respectively. The spin
operator $\sigma^{ij}_{-\mu}$ destroys or creates quark-antiquark pairs with spin 1.

The internal spatial wave functions for both the baryons and mesons are taken in the
harmonic oscillator approximation in the work. The $S$-wave meson wave function can be expressed
in terms of the quark
momenta as
\begin{eqnarray}
\langle M |\vec{q}_{i}
\vec{q}_{j}\rangle
=N_M\,{\rm
exp}\left\{-\frac{b^2}{8} \Big(\vec{q}_{i} -\vec{q}_{j}\Big)^2
\right\}\chi_{M},
\end{eqnarray}
with $N_M = (b^2/\pi)^{3/4}$ and $b$ is the meson radial
parameter. The spin-color-flavor wave function is denoted by
$\chi_M$. The baryon wave functions are given by
\begin{eqnarray}
&& \langle B|\vec{q}_i \vec{q}_j \vec{q}_k\rangle \nonumber \\
&=& N_B\,{\rm exp}\left\{-\frac{a^2}{4}
\Big[(\vec{q}_j-\vec{q}_k)^2
+\frac{(\vec{q}_j+\vec{q}_k-2\vec{q}_i)^2}{3}\Big]  \right\}
\chi_{B}, \nonumber \\
\end{eqnarray}
where $N_B=(3a^2/\pi)^{3/2}$ with $a$ being the baryon radial
parameter, and $\chi_{B}$ is the spin-color-flavor wave function of
baryons.

The transition amplitude of $N\overline N$ annihilations to two mesons
can be written in terms of the partial waves of the initial and final states,
\begin{eqnarray}\label{eq::6}
T_{\overline NN \to M_1M_2}(\,\vec p,\vec k\,) &=& \sum_{LM}\sum_{lm}Y_{LM}(\hat p)\,T_{fi}(p,k)\,Y_{lm}^*(\hat k)
\nonumber \\
\end{eqnarray}
In the low-momentum approximation, the partial wave transition amplitude $T_{fi}(p,k)$
is derived in the A2 quark line diagram as
\begin{eqnarray}\label{eq::7}
T_{fi}(p,k)=\lambda^3\,F_{l}^L(p)\,G_{l}^L(k) \langle f | O_{A_2}|i \rangle \nonumber \\
\cdot{\rm exp} \left\{ -Q^2_p\,p^2 -Q^2_k\,k^2\right\}
\end{eqnarray}
The index $i$ represents the initial state $^{2I+1,2S+1} L_J$ with
$I$, $J$, $L$ and $S$ being respectively the total isospin, total angular momentum,
orbital angular momentum and total spin while the index $f$ stands
for the final two meson states with $l$ being the relative orbital angular momentum between
the final two mesons. $F_{l}^L(p)$ is a function of $p=|\vec p|$ and the meson and baryon
radial parameters $a$ and $b$ while $G_{l}^L(k)$ is a function of $k=|\vec k|$, $a$ and $b$.
$Q^2_p$ and $Q^2_k$ are geometrical constants
depending on the radial parameters. The matrix elements $\langle f |
O_{A_2}|i \rangle$ are the spin-flavor weights for the quark line
diagram $A_2$. Listed in Table I are the $\langle f |
O_{A_2}|i \rangle$ values, normalized to the $\rho\rho$ channel,
for annihilations of the initial $^{11}S_0$ $N\overline N$ state to various two-meson channels.

In the low-momentum approximation, $G^L_l(k)$ in eq. (\ref{eq::7}) are derived as
\begin{eqnarray}\label{eq::8}
G_{l=1}^{L=0}(k) &=& k\,(1+A\,k^2)
\end{eqnarray}
with
\begin{eqnarray}\label{eq::9}
A &=& -\frac{ a^2 b^4 \left(3 a^2+b^2\right)}{2 \left(3 a^2+2
b^2\right) \left(9 a^4+13 b^2 a^2+3 b^4\right)}
\end{eqnarray}
for
the annihilation processes of the initial $^{11}S_0$ $N\overline N$ state to
two $s$-wave mesons, and
\begin{eqnarray}\label{eq::10}
G^{L=0}_{l=0}(k) &=& B_1\,(1+B\,k^2)\nonumber \\
G^{L=0}_{l=2}(k) &=& \,C_1\,k^2
\end{eqnarray}
for
the reactions of the initial $^{11}S_0$ $N\overline N$ state to
the final states of one $s$-wave and one $p$-wave mesons,
where $G^{L=0}_{l=0}(k)$ ($G^{L=0}_{l=2}(k)$) is for the final state with one $p$-wave meson of spin $J=0$ ($J=2$).
$B$, $B_1$ and $C_1$ in eq. (\ref{eq::10})
are functions of the size parameters $a$ and $b$, taking the forms
\begin{eqnarray}
B &=& \frac{a^2 \left(27 a^4+45 b^2 a^2+8 b^4\right) }{6 \left(6 a^4+13 b^2 a^2+6 b^4\right)}, \nonumber \\
B_1 &=& \frac{b \left(2 a^2+3 b^2\right)}{9 a^4+13 b^2 a^2+3 b^4}, \nonumber \\
C_1 &=& \frac{ a^2 b \left(27 a^4+45 b^2 a^2+11 b^4\right)}{27 a^6+57 b^2 a^4+35 b^4 a^2+6 b^6}.
\end{eqnarray}
Note that we have let $F_{l}^L(p)$ in eq. (\ref{eq::7}) the same for all annihilation channels just for convenience.
\begin{table}
\begin{center}
\label{table::1} \caption{$\langle f |
O_{A_2}|i \rangle$ and partial decay widths
for annihilations of the initial $^{11}S_0$ $N\overline N$ state to two mesons.} \vspace*{.3cm}
\begin{tabular}{|c|c|c|}
\hline
Final states & $\langle f |
O_{A_2}|i \rangle$ & $\Gamma_i/\Gamma_{X(1835)\to\rho\rho}$\\
\hline
 $\rho\rho$ & $1$ & 1 \\
\hline
 $\omega\omega$ & $1/\sqrt{3}$  & 0.32 \\
\hline
 $\pi a_0(1450)$ & $\sqrt{2}$  & $0.25 $ \\
\hline
 $\eta f_0(1370)$ & $1$ & $ 0.057 $ \\
\hline
 $\eta' f_0(1370)$ & $1$ & $0.052 $ \\
\hline
 $\pi a_2(1320)$ &  $1/\sqrt{6}$ & 0.055 \\
\hline
 $\eta f_2(1270)$ & $1/(3\sqrt{2})$ & 0.008\\
\hline
\end{tabular}
\end{center}
\end{table}

The transition amplitude of annihilations of $N\overline N$ bound states
takes the form,
\begin{eqnarray}\label{atomic state}
T_{f,LSJ}^I(k)=\int p^2dp\, T_{fi}(p,k) \psi^I_{LSJ}(p),
\end{eqnarray}
where $ \psi^I_{LSJ}(p)$ is the radial wave function of the initial $N\overline N$ bound state in
momentum space. The partial decay width for
the transition of $N\overline N$ bound states to two mesons is given by
\begin{eqnarray}
\Gamma_{p\bar{p}\rightarrow M_1M_2} &=& \frac{1}{2M}\int\frac{d^3k_1}{2E_1}\frac{d^3k_2}{2E_2}\delta ^{(3)}(\vec{k}_1 +\vec{k}_2) \nonumber \\
&&\delta(E-E_1-E_2 )|T_{f,LSJ}(\vec{k})|^2
\end{eqnarray}
where $M$ is the mass of the $N\overline N$ bound state, and $E_{1 ,2}=
\sqrt{m^2_{1,2}+\vec{k}^2_{1,2}} $ is the energy of outgoing
mesons with mass $m_{1,2}$ and momentum
$\vec{k}_{1,2}$. With the explicit form of the transition
amplitude given by Eq. (\ref{eq::7}), the partial decay width
for the $X(1835)$ annihilation to two mesons is derived as
\begin{eqnarray}\label{decaywidth}
\Gamma_{X \rightarrow M_1M_2}=\lambda_{A_2}^2\langle f | O_{A_2}|i \rangle ^2 |G^L_l(k)|^2\,F_p F_k ,
\end{eqnarray}
with
\begin{eqnarray}
F_p= |F_{l}^{L}(p)  \int p^2dp ~\psi^I_{LSJ}(p) { \rm exp} \left\{ -Q_{p}^2\, p^2\right\}|^2
\end{eqnarray}
and the kinematical phase-space factor defined by
\begin{equation}\label{phasespacefactor}
F_k=\frac{k}{8M^2} {\rm exp} \left\{ -2Q_{k}^2\, k^2 \right\}.
\end{equation}
The spin-flavor weights $\langle f | O_{A_2}|i \rangle$ are listed in Table I
for the transition $X(1835)\to M_1M_2$.

The model dependence in Eq.(\ref{decaywidth}) may be reduced by
choosing a simplified phenomenological approach that has
been applied in studies of two-meson branching ratios in
nucleon-antinucleon annihilation \cite{Kercek:1999sc} and radiative protonium annihi-
lation \cite{Gutsche:1998fc}. Namely, instead of the phase space factor of
Eq. (\ref{phasespacefactor}) obtained in the harmonic oscillator
approximation for the
hadron wave function, we use
a kinematical phase-space factor of the phenomenological
form
\begin{eqnarray}\label{f-function}
f(\phi,X)=k\cdot {\rm exp}\{-a_s\,(s-s_{12})^{1/2}\}
\end{eqnarray}
where $s_{12}=(m_{1}+m_{2})^{2}$ and
$\sqrt{s}=(m_{1}^2+k^2)^{1/2}+(m_{2}^2+k^2)^{1/2} $. The constant
$a_s=1.2$ GeV$^{-1}$ is obtained from the fit to the momentum dependence
of the cross section of various $\overline NN$
annihilation channels \cite{Vandermeulen:1988hh}.

Partial decay widths are evaluated for the
processes of a $^{11}S_0$ state $X(1835)$ to $\rho\rho$, $\omega\omega$,
$\pi a_0(1450)$, $\eta f_0(1370)$, $\eta' f_0(1370)$,
$\pi a_2(1320)$ and $\eta f_2(1270)$. Here the $a_0(1450)$, $f_0(1370)$,
$a_2(1320)$ and $f_2(1270)$ are normal $p$-wave mesons, and their radial wave functions take the Gaussian form.
It is rather difficult to calculate, in the scope of this work and also any quark model, the partial
decay widths to the states $\eta f_0(600)$, $\eta' f_0(600)$ and $\pi a_0(980)$ as the nature of the
$f_0(600)$ and $a_0(980)$ mesons is not clear. As the $a_0(980)$ and $f_0(600)$ mesons
may have large non-$\overline qq$ components \cite{Jaffe,Bugg,Tornqvist}, we may expect that the decays of
the $X(1835)$ to
the channels $\pi a_0(980)$, $\eta f_0(600)$ and $\eta' f_0(600)$ are less important.

The theoretical results of partial decay widths are very sensitive to
the effective coupling constants
$\lambda$ of the $^3P_0$ vertex and the size parameters $a$ and $b$.
The meson size parameter $b$ is determined to be 3.24 GeV$^{-1}$ by
the reaction $\rho \to e^+e^-$ as in Refs.
\cite{Yan:2004jg,Kittimanapun:2008wg}, which leads to an rms radius $\langle r^2\rangle^{1/2}=0.39$ fm for the
$s$-wave mesons, while the optimum meson size parameter derived by
fitting to the partial widths of higher quarkonia \cite{Barnes:1996ff} is
$b=2.5$ GeV$^{-1}$. However, in studies of $\overline
NN$ annihilations \cite{Maruyama:1987tx,Maruyama:1987kt,Dover:1990,Gutsche:1989qk,Muhm:1996tx,Dover:1992vj},
the meson size parameter $b$ is
globally adjusted
to 4.1 GeV$^{-1}$ ($\langle r^2\rangle^{1/2}=0.50$ fm).
As for the baryon size parameter $a$, various values
have been employed in different works,
ranging from $a=1.6$ GeV$^{-1}$ to $3.1$
GeV$^{-1}$ \cite{Maruyama:1987tx,Maruyama:1987kt,Dover:1990,Gutsche:1989qk,Muhm:1996tx,
Dover:1992vj,Mundigl:1991jp,Isgur:1978xj,Isgur:1979be,Capstick:1986bm,Chen:2007xf,Capstick:1992th,Capstick:1993kb}.

Theoretical results of partial decay widths are expected to be sensitive to the
$N\overline N$ bound state wave function $\Phi_X(\vec p)$ of the $X(1835)$. However,
our poor knowledge of $\overline NN$ interactions does not allow us to work out
a reliable wave function $\Phi_X(\vec p)$.
Therefore, we evaluate in this work the relative partial decay widths to avoid the uncertainties
that we have rather poor knowledge of the effective
coupling constant $\lambda$ and the $N\overline N$ bound state wave function $\Phi_X(\vec p)$ of the $X(1835)$,
and that the meson and nucleon size parameters $a$ and $b$ may range in rather large regions.

It is found that the ratios of partial decay widths are, of course,  independent of the effective coupling
strength $\lambda$ of the $^3P_0$ vertex, insensitive to the $N\overline N$ bound state wave function $\Phi_X(\vec p)$,
and insensitive to the meson and baryon size parameters. Indeed, in the low momentum approximation,
the ratios of partial decay widths are independent of the the $N\overline N$ bound state wave
function $\Phi_X(\vec p)$, as shown in Eq. (\ref{decaywidth}).
As an example, we
shown in Table I the relative partial decay widths, normalized to the $\rho\rho$ channel, of a $^{11}S_0$ state $X(1835)$
to two mesons,
with the baryon size parameter $a=2.0$ GeV$^{-1}$ and meson size parameter
$b=3.24$ GeV$^{-1}$. In the calculation the $\eta$ and $\eta'$ mesons
are represented as
\begin{eqnarray}\label{mixture1}
|\eta'\rangle &=& \beta\,|\eta_n\rangle+\alpha\,|\eta_s\rangle \nonumber\\
|\eta\rangle &=& \alpha\,|\eta_n\rangle-\beta\,|\eta_s\rangle
\end{eqnarray}
in the basis
\begin{eqnarray}
|\eta_n\rangle &=& \frac{1}{\sqrt{2}}\left(|u\overline u\rangle+|d\overline d\rangle \right), \nonumber \\
|\eta_s\rangle &=& |s \overline s\rangle,
\end{eqnarray}
where $\alpha$ and $\beta$ are given in terms of the pseudoscalar
mixing angle $\theta$ by the relation
\begin{eqnarray}
\alpha &=& \sqrt{\frac{1}{3}}\cos\theta - \sqrt{\frac{2}{3}}\sin\theta, \nonumber \\
\beta &=& \sqrt{\frac{1}{3}}\cos\theta +
\sqrt{\frac{2}{3}}\sin\theta
\end{eqnarray}
We take the canonical value $\theta=-10.7^o$ derived from the
quadratic mass formula, which leads to $\alpha\approx\beta\approx 1/\sqrt{2}$.

For the $X(1835)$ and broad mesons we average over the mass spectrum $f(\mu)$, that is
\begin{eqnarray}
\overline\Gamma &=& \int d\mu_X f_X(\mu_X)\int d\mu_1 f_1(\mu_1)\int d\mu_2 f_2(\mu_2)\nonumber \\
&& \Gamma_{X \rightarrow M_1M_2}(\mu_X,\mu_1,\mu_2)
\end{eqnarray}
with the mass spectrum $f(\mu)$ as in Ref. \cite{Dover:1992vj}
\begin{eqnarray}
f_i(\mu)=C\frac{(\Gamma_i/2)^2}{(\mu-M_i)^2+(\Gamma_i/2)^2}
\end{eqnarray}
where $C$ is a normalization constant, and $\Gamma_{X \rightarrow M_1M_2}$ are derived in eq. (\ref{decaywidth}).

One sees in Table I that the $\rho\rho$, $\omega\omega$ and $\pi a_0(1450)$ decay channels dominate over others. As
the branching fractions of the $J/\psi$ one photon radiative decays are in order of $10^{-4}$ to $10^{-3}$ \cite{particledata},
the branching fractions
$Br(J/\psi\to\gamma X)Br(X\to\rho\rho,\;\omega\omega,\;\pi a_0(1450))$ are expected to be in order of
$10^{-5}$ to $10^{-3}$. In high energy $e^+e^-$ collisions $\pi$ mesons are produced in a dominant number,
which may make it difficult
to retrieve the $\rho\rho$ and $\omega\omega$ channels. However, one may expect that the resonance $X(1835)$ is observed in the
$\pi a_0(1450)$ channel given it is a $\overline NN$ bound state.

\section{$X(1835)$ to $\eta\pi\pi$ and $\eta'\pi\pi$}
The $X(1835)$ resonance is observed in the $\eta'\pi\pi$ channel with the product branching fraction
$Br(J/\psi\to\gamma X)Br(\to\pi^+\pi^-\eta') = (2.2\pm 0.4({\rm stat.})\pm 0.4({\rm syst.}))\times 10^{-4}$.
In this section we estimate the joint contribution of the $\pi a_0(1450)$ and $f_0(1370)\eta$ channels
to the final states $\pi\pi\eta$ and $\pi\pi\eta'$. The broad meson $f_0(1370)$
may decay to $2\pi$, $4\pi$, $\eta\eta$ and
$K\overline K$, with the $2\pi$ channel predominant. The refit in Ref. \cite{Bugg2} to five primary sets
of data leads to the partial decay widths,
\begin{eqnarray}
\Gamma_{f_0(1370)\to\pi\pi} &=& 325\, {\rm MeV}, \nonumber \\
\Gamma_{f_0(1370)\to 4\pi} &=& 54\, {\rm MeV}, \nonumber \\
\frac{\Gamma_{f_0(1370)\to \eta\eta}}{\Gamma_{f_0(1370)\to \pi\pi}} &=& 0.19 \pm 0.07.
\end{eqnarray}
Considering that the $K\overline K$ decay channel is usually strongly suppressed, one may
estimate
$\Gamma_{f_0(1370)\to \pi^+\pi^-}/\Gamma_{\rm tol}\sim 0.5$
and hence
\begin{eqnarray}
\frac{\Gamma(X\to f_0(1370)\eta\to \eta\pi^+\pi^-)}{\Gamma(X\to \rho\rho)}\sim 0.029 \nonumber \\
\frac{\Gamma(X\to f_0(1370)\eta'\to \eta'\pi^+\pi^-)}{\Gamma(X\to \rho\rho)}\sim 0.026
\end{eqnarray}
The $a_0(1450)$ meson decays dominantly to $\pi\eta$, $\pi\eta'$ and $K \overline K$ channels,
with the experimental values
\begin{eqnarray}
\frac{\Gamma_{a_0(1450)\to \eta'\pi}}{\Gamma_{a_0(1450)\to \eta\pi}}\,=\, 0.35 \pm 0.16,\nonumber \\
\frac{\Gamma_{a_0(1450)\to K\overline K}}{\Gamma_{a_0(1450)\to \eta\pi}}\,=\, 0.88 \pm 0.23.
\end{eqnarray}
We may estimate the contributions of the $\pi a_0(1450)$ intermediate channel to the final states
$\eta\pi^+\pi^-$ and $\eta'\pi^+\pi^-$,
\begin{eqnarray}
\frac{\Gamma(X\to a_0(1450)\pi\to \eta\pi^+\pi^-)}{\Gamma(X\to \rho\rho)}\sim 0.075 \nonumber \\
\frac{\Gamma(X\to a_0(1450)\pi\to \eta'\pi^+\pi^-)}{\Gamma(X\to \rho\rho)}\sim 0.025
\end{eqnarray}
Assuming that there is no interference between the contributions of the $\pi a_0(1450)$ and $f_0(1370)\eta$
intermediate channels, one may derive
\begin{eqnarray}\label{eq::26}
\frac{\Gamma(X\to \eta\pi^+\pi^-)}{\Gamma(X\to \rho\rho)}\sim 0.1 \nonumber \\
\frac{\Gamma(X\to \eta'\pi^+\pi^-)}{\Gamma(X\to \rho\rho)}\sim 0.05
\end{eqnarray}

Given that
the branching ratio $Br(J/\psi\to\gamma X(1835))$ is in order of $10^{-4}$ to $10^{-3}$ \cite{particledata},
as other one photon $J/\psi$
radiative decays, the theoretical estimations in eq. (\ref{eq::26})
indicates that the product branching fraction
$Br(J/\psi\to\gamma X(1835))Br(X(1835)\to \pi\pi\eta')$ is in order of $10^{-5}$ to $10^{-4}$,
which is line with the experimental data.

\section{Discussion and Conclusions}
Partial decay widths of the $X(1835)$ to various decay channels are evaluated in the $^3P_0$ quark model,
assuming that the $X(1835)$ is a $N\overline N$ bound state with the quantum number
assignment $I^G(J^{PC})=0^+(0^{-+})$. We find that the decays of the $X(1835)$ to $\rho\rho$, $\omega\omega$ and $\pi a_0(1450)$
dominate over other channels. Based on the large $\pi a_0(1450)$ partial decay width, we would like to suggest the resonance
$X(1835)$ to be searched in the $\pi a_0(1450)$ channel.

The contributions of the $\pi a_0(1450)$, $\eta f_0(1370)$ and $\eta' f_0(1370)$ channels to the
final states $\pi\pi\eta$ and $\pi\pi\eta'$ are estimated. It is found that the partial decay widths
$\Gamma(X\to \eta\pi^+\pi^-)$ and $\Gamma(X\to \eta'\pi^+\pi^-)$ are in the same order.
The product branching fraction $Br(J/\psi\to\gamma X)Br(X\to \pi\pi\eta')$
is estimated to be in order of $10^{-5}$ to $10^{-4}$,
which is line with the experimental data.

As a pseudoscalar meson, the $X(1835)$ may decay through the $KK^*$ and $K^*K^*$ channels. However, these decay modes
of a $\overline NN$ bound state are strongly suppressed \cite{Amsler1991,Amsler1993}. It is expected that the
product branching fraction $Br(J/\psi\to\gamma X)Br(X\to KK^*,\;K^*K^*)$ is even smaller than
$Br(J/\psi\to\gamma X)Br(X\to \pi\pi\eta')$ if the $X(1835)$ is interpreted as an $\overline NN$
bound state.

It is natural to interpret the $X(1835)$ as the second radial excitation of $\eta'$ as it is
observed in the $\eta'\pi\pi$ channel. A quark model study \cite{etap4} shows that the $X(1835)$ decays to
$KK^*$ and $K^*K^*$ states with large partial decay widths. The
product branching fraction $Br(J/\psi\to\gamma X)Br(X\to KK^*,\;K^*K^*)$ shall be much larger than
$Br(J/\psi\to\gamma X)Br(X\to \pi\pi\eta')$.
Whether the $X(1835)$ could be observed in the $KK^*$ and $K^*K^*$ channels
with large product branching fractions may tell
if it is a radial excitation of $\eta'$.

\section*{Acknowledgements}
The work was partly supported by the Commission on Higher Education, Thailand with
the grant CHE-PhD-SW-SUPV.

\end{document}